\documentclass{elsart}

\usepackage{graphicx}

\usepackage{amssymb}

\begin{document}

\begin{frontmatter}

\title{New aspects on Current enhancement in Brownian motors
driven by non Gaussian noises}

\author[label1]{Sebasti\'an Bouzat}
\author[label1,label2]{Horacio S. Wio}

\address[label1]{Centro At\'omico Bariloche (CNEA) and Instituto
Balseiro (UNC-CNEA)\\ 8400 San Carlos de Bariloche, R\'{\i}o
Negro, Argentina}

\address[label2]{Instituto de F\'{\i}sica de Cantabria, Avda. Los
Castros s/n \\ E-39005 Santander, Spain}

\begin{abstract}
Recent studies on Brownian motors driven by colored non Gaussian
noises have shown that the departure of the noise distribution
from Gaussian behavior induces an enhancement of its current and
efficiency. Here we discuss some new aspects of this phenomenon
focusing in some analytical results based in an adiabatic
approximation, and in the analysis of the long probability
distribution tails' role.
\end{abstract}

\end{frontmatter}

\section{Introduction}

Recent studies on the role of non Gaussian noises on several
``noise-induced phenomena", like stochastic resonance, resonant
trapping, and noise-induced transitions
\cite{qRE1,qRE1-2,qRE1p,qRE2,qruido1,qruido2,qruido3}, have shown
the existence of strong effects on the system's response. The form
of the noise source used was based on the nonextensive statistics
\cite{tsallis,tsallis1,tsallis2} with a probability distribution
that depends on $q$, a parameter indicating the departure from
Gaussian behavior: for $q = 1$ the distribution is Gaussian, while
different non Gaussian distributions result for $q > 1$ or $q <
1$. What was observed was an enhancement of the signal-to-noise
ratio in stochastic resonance, an enhancement of the trapping
current in resonant trapping, and a marked shift in the transition
line of noise-induced transitions.

Those studies motivated us to also analyze the effect of non
Gaussian noises on the behavior and transport properties of
Brownian motors \cite{qruidoZ}. As is well known the study of
noise induced transport by ``ratchets" has attracted in recent
years the attention of an increasing number of researchers due to
its biological interest and also to its potential technological
applications \cite{review,review1,review2,review3}.

In \cite{qruidoZ} we have shown that, under certain general
conditions, non Gaussian noise induces current and efficiency
enhancement, without the need of a fine tuning of the parameters.
Also we have found the phenomenon of current inversion as the
parameter $q$ (associated to the non Gaussian properties of the
noise) is varied. The analysis of the effects of a colored non
Gaussian noise source on the transport properties of a Brownian
motor was done by considering two alternative points of view: one
which can be interpreted as more directly connected to
technological applications and the other more related to
biological or natural systems \cite{qruidoZ}. In the model
analyzed, these two visions correspond to consider as control
parameters, different (but related) constants that are associated
to correlation times and noise amplitudes. In this work we will
focus on the second point of view, which considers the non
Gaussian noise as the primary or direct forcing of the Brownian
motor.

In \cite{qruidoZ} we have analyzed the system considering a long
correlation time and a small amplitude of the stochastic forcing.
In this contribution we extend the results found in \cite{qruidoZ}
by relaxing the condition of small amplitude of the noise. In this
way we find new regimes, where the dynamics of the Brownian motor
is affected in different ways by the non Gaussianity of the noise.
We also delve deeper into the analytical study of the system in
the adiabatic approximation (which assumes a large correlation
time of the forcing) and we give a more complete understanding of
some of the phenomena studied in \cite{qruidoZ}. Finally, we
analyze the role played by the long tails of the distribution, and
compare the results of truncated non Gaussian distributions for
different cutoff values but having the same width. In this way we
show the crucial role played by those long tails in determining
the enhancement of the Brownian motor response.

\section{Model and approximations}

We consider the following model for a particle in a rocked
ratchet, which is the same studied in Ref. \cite{qruidoZ} but
considering the overdamped limit
\begin{equation}
\label{xpto} \frac{dx}{dt}=-V'(x)-F+\xi(t)+\eta(t),
\end{equation}
where $V(x)$ is the ratchet potential, $F$ is a constant ``load''
force, and $\xi(t)$ the thermal noise satisfying $\langle \xi(t)
\xi(t') \rangle= 2 T \delta(t-t')$. Finally, $\eta(t)$ is the time
correlated forcing (with zero mean) that allows the rectification
of the motion, keeping the system out of thermal equilibrium even
for $F=0$. For the ratchet potential we consider the same form as
in \cite{magnasco} and \cite{qruidoZ} (for instance, see Fig. 3 in
\cite{qruidoZ})
\begin{equation}
\label{V1dex} V(x)=V_1(x)=- \int^x dx' \left( \frac{\exp[\alpha
\cos(x')]}{J_0(i\alpha)} - 1 \right),
\end{equation}
where $J_0(i\alpha)$ is the Bessel function, and $\alpha=16$.

We will consider the dynamics of $\eta(t)$ as described by the
following Langevin equation \cite{qRE1,qRE1-2}
\begin{equation}
\label{etapto} \frac{d\eta}{dt}=-\frac{1}{\tau}\frac{d}{d \eta}
V_q(\eta) + \frac{1}{\tau} \zeta(t),
\end{equation}
with $\langle\zeta(t)\rangle=0$ and
$\langle\zeta(t)\zeta(t')\rangle=2D\delta(t-t')$, and
$$V_q(\eta) = \frac{D}{\tau(q-1)}\,
\ln[1+\frac{\tau}{D}(q-1)\frac{\eta^2}{2}].$$

For $q=1$, the process $\eta$ coincides with the
Ornstein-Uhlenbeck (OU) one (with a correlation time equal to
$\tau$), while for $q \neq 1$ it is a non Gaussian process. As
shown in \cite{qRE1,qRE1-2}, the form of the probability
distribution function (pdf), for $q$ within the range $-\infty < q
< 3$, is
\begin{equation}\label{pdfq}
P_q(\eta)=\frac{1}{Z_q} \left[1+\frac{\tau (q-1)\eta^2}{2
D}\right]^\frac{1}{1-q},
\end{equation}
that for $1<q<3$, extents along the interval $-\infty < \eta <\infty$,
and decays as a power law (slower than a Gaussian
distribution). For $q<1$, the sign inside the brackets is changed,
and the pdf has a bounded support with a cut-off at $|\eta|=
\omega \equiv [(1-q)\tau/(2D)]^{-\frac{1}{2}}$. $Z_q$ is a
normalization constant. For $q>3$ this distribution can not be
normalized.

The main characteristic introduced by this non Gaussian form of
the forcing is that, for $q>1$, the distribution decays as a power
law. This leads to the appearance of arbitrary strong ``kicks'' on
the ratchet particle with relatively high probability when
compared, for example, with the Gaussian OU noise and, of course,
with the dichotomic non Gaussian process. For a picture of the
typical form of this pdf for different values of $q$, we refer to
Fig. 1 in \cite{qruidoZ}.

In \cite{qRE1,qRE1-2} it was shown that the second moment of the
distribution, which we interpret as the ``intensity" of the non
Gaussian noise, which is given by
\begin{equation}\label{eqn5}
D_{ng} \equiv \langle \eta^2 \rangle =\frac{2D}{\tau (5-3q)},
\end{equation}
diverges for $q \ge 5/3$. For the correlation time $\tau_{ng}$ of
the process $\eta(t)$, defined in detail in \cite{qRE1,qRE1-2} it
was not possible to find an analytical expression. However, it is
known \cite{qRE1,qRE1-2} that for $q \to 5/3$ it diverges as $\sim
(5-3q)^{-1}$. In our analysis, we will consider values of $q$ in
the range $0.5<q<5/3\simeq1.66$. For this interval we have studied
numerically the dependence of $\tau_{ng}$ on $q$, and have found
the following analytical approximation
\begin{equation}
\label{fittng}  \tau_{ng} \simeq 2 \frac{[1+ 4 (q-1)^{2}] \,
\tau}{(5-3q)},
\end{equation}
that fits very accurately the results. This fitting will be the
one we will consider when analyzing the dependence of the
transport properties of the ratchet system on the intrinsic
parameters of the non Gaussian noise, $D_{ng}$ and $\tau_{ng}$.

As mentioned in the introduction, in \cite{qruidoZ} we have
studied the current and efficiency of the system as function of
$q$ from two points of view. One corresponds to consider $D$ and
$\tau$ as control parameters (in addition to $q$) and the other is
to consider $D_{ng}$ and $\tau_{ng}$ instead. The first point of
view is the more direct one when thinking on tailor made
technological devices \cite{electr,electr1,electr2}. The second
one is the more natural from the point of view of biological
systems, as it consider $\eta$ as a primary source of noise
characterized by its intensity and correlation time. As indicated
in the introduction, in this work we focus on this second view.

Considering the first point of view, in \cite{qruidoZ} we have
found that a departure from Gaussian behavior (particularly for
$q>1$), induces a remarkable increase of the current together with
an enhancement of the motor efficiency. The efficiency shows in
addition an optimum value for a given degree of departure from the
Gaussian behavior and decays due to the enhancement of
fluctuations when the correlation of the non Gaussian noise
diverges. When inertia is taken into account we have also found a
considerable increment in the mass separation capability of the
system.

The main results in \cite{qruidoZ} corresponding to the second
point of view will be discussed in the following sections in
connection with the new results we will show. In the next section,
we extend the analysis made in \cite{qruidoZ} -- which involves
only relatively low intensities of the non Gaussian noise --
providing analytical results for a wide range of $D_{ng}$.
Different regions of this parameter are found where $q_o$, the
optimum value of $q$ that maximizes the current, changes from
$q_o>1$ to $q_o<1$.

\section{Analytical results within the adiabatic approximation.}

In the overdamped regime we are able to give an approximate
analytical solution for the problem, which is valid in the large
correlation time regime ($\frac{\tau}{D} \gg 1$): we perform the
adiabatic approximation of solving the Fokker-Planck equation
associated to Eq. (\ref{xpto}) assuming a constant value of $\eta$
\cite{comFP}, see e.g. \cite{qruidoZ}. This leads us to obtain an
$\eta$--dependent value of the current $J(\eta )$ that is then
averaged over $\eta$ using the distribution $P_q(\eta)$ in Eq.
(\ref{pdfq}) \cite{qruidoZ} with the desired values of $q$, $D$
and $\tau$
$$J = \int d\eta \, J(\eta ) \, P_q(\eta).$$
We remark that, although the Fokker-Planck equation is solved in
the $\tau/D \to \infty$ limit, the solution we find depends on $D$
and $\tau$ through the $P_q(\eta)$ distribution. Note that, in
order to perform the adiabatic approximation, it is essential to
consider a non vanishing temperature, since this gives the random
ingredient that leads to a Fokker-Planck equation. In Fig. 1 we
show a typical curve for $J(\eta)$.

\begin{figure}
\centering
\includegraphics[width=9cm]{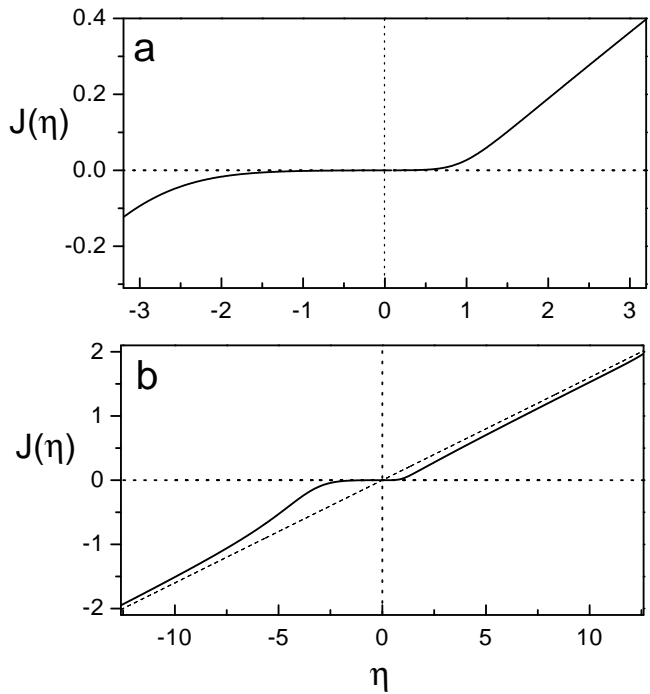}
\caption{Results for $J(\eta)$ for $F=0.1$ and $T=0.5$. In both
parts of the figure the same curve is plotted at different scales
in order to better show the behavior. In part (b) we have also
shown the line to which $J(\eta)$ converges for $\eta \to \pm
\infty$.}
\end{figure}

Now we study the dependence of the current $J$ on the parameters
$D_{ng}$ and $q$. We consider the range $0<D_{ng}<200$ and
$0.5<q<1.66$. (For $D_{ng}>100$ the current decays as a
consequence of the ``excess" of noise, as is typical in most of
the noise induced phenomena. Hence, the behavior of the system for
higher values of $D_{ng}$ is not considered to be relevant.) For
the rest of the parameters we consider values similar to those in
\cite{magnasco} and \cite{qruidoZ} where interesting transport
phenomena have been observed. We fix $T=0.5, F=0.1$ and
$\tau_{ng}=100/(2\pi) \sim 15.9$. The relevance of this parameter
region was discussed in \cite{qruidoZ}. As we show in the next
section, the value set for $\tau_{ng}$ is high enough to make the
adiabatic approximation valid.

In Fig. 2 we show results for the averaged current $J$ as function
of $D_{ng}$ for different values of $q$. Parts (a) and (b) of the
figure show the same curves on different scales in order to better
appreciate the crossings' details. At a first glance, one may
observe that for $D_{ng} \lesssim 0.4$ the current increases with
$q$, for $0.4 \lesssim D_{ng} \lesssim 30$ the current decreases
with $q$, and finally, for $D_{ng} > 30$ the current increases
with $q$ again. However, what actually occurs is a little more
complicated. The results should be carefully read, as we are not
showing the curves for all the possible values of $q$. In
\cite{qruidoZ} we have analyzed the behavior of the system in the
region of small $D_{ng}$. It was shown that, for a fixed value of
$D_{ng} < 0.4$, there is an optimum value of $q>1$, $q_o$, that
maximizes the current. What happens in Fig. 2a for very low values
of $D_{ng}$ is that $q_o$ is larger than the highest value plotted
for $q$. However, approximately at $D_{ng}=0.1$ a crossing between
the curves for $q=1.55$ and $q=1.4$ occurs. This means that the
optimum value became lower than $q=1.55$.

\begin{figure}
\centering
\includegraphics[width=12cm]{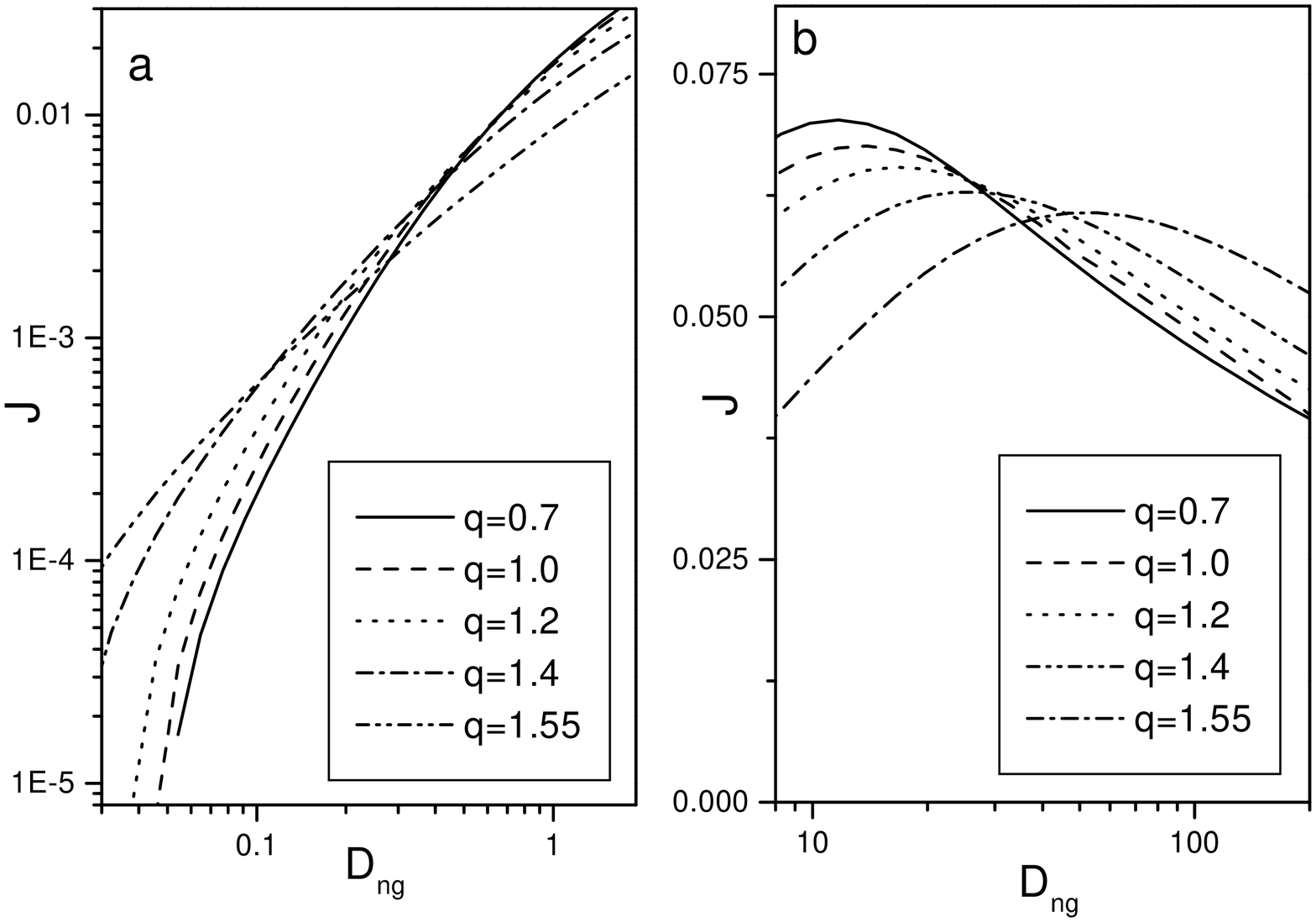}
\caption{Current vs $D_{ng}$ for different values of $q$,
calculated within the adiabatic approximation. Both parts of the
figure show the same curves in different ranges of $D_{ng}$.}
\end{figure}

Now, considering that, as discussed in \cite{qruidoZ}, the current
should vanish for very low values of $q$, since the bounds for the
distribution of $\eta$ are reduced, and that for $q \to 5/3$ the
current decreases due to the increasing of the fluctuations of
$\eta$, an optimum value of $q$ should always be expected. Hence,
the results in Fig. 2 should be interpreted as follows. For every
value of $D_{ng}$ there is an optimum value of $q \equiv q_o$ that
maximizes the current. For $D_{ng} \lesssim 0.4$ we have $q_o>1$,
for $0.4 \lesssim D_{ng} \lesssim 30$ we have $q_o<1$, and for $40
< D_{ng}$ we have again $q_o>1$.

In the following section we present results from numerical
simulations that validate these predictions of the adiabatic
approximation.

\section{Numerical simulations}

We have analyzed numerically the evolution of Eqs. (\ref{xpto})
and (\ref{etapto}) by numerical integration of those equations
using a second order stochastic Runge--Kutta type algorithm
\cite{sanmiguel}. The current is defined as $J=\frac{\langle
\dot{x} \rangle}{L}$ where $L=2 \pi$ is the period of the ratchet
potential, and $\langle \rangle$ indicates temporal averaging.

Here, in order to appreciate the maxima predicted in the previous
section, we present results for the current as function of $q$ for
different values of $D_{ng}$. We have only considered values of
$D_{ng}\le 20$, as higher values of this parameter requires
considerable computational effort. In Fig. 3 we show the results
coming from simulations together with the curves obtained
analytically from the adiabatic theory. It can be seen that, for
the value of $\tau_{ng}$ considered, concerning the general
behavior of the results as function of $q$ and $D_{ng}$ there is a
rather good agreement between theory and simulations. Better
accuracy from the adiabatic theory can only be obtained for larger
values of the correlation time.

The results in Fig 3. confirm the predictions of the previous
section (at least for the values of $D_{ng} \leq 30$ here
considered), as the optimum value of $q$ appears to be $q_o>1$ for
$D_{ng} \lesssim 0.4$ and $q_o<1$ for $0.4 \lesssim D_{ng}
\lesssim 30$. It should be mentioned that some of the results on
this figure (those for $D_{ng}<1$) have been presented in
\cite{qruidoZ}. However, we want to remark that in that work we
have only explored the region of low $D_{ng}$ while now, we have a
more panoramic view of the system's behavior as function of that
parameter that ranges from $0$ to $\infty$.

\begin{figure}
\centering
\includegraphics[width=10.5cm]{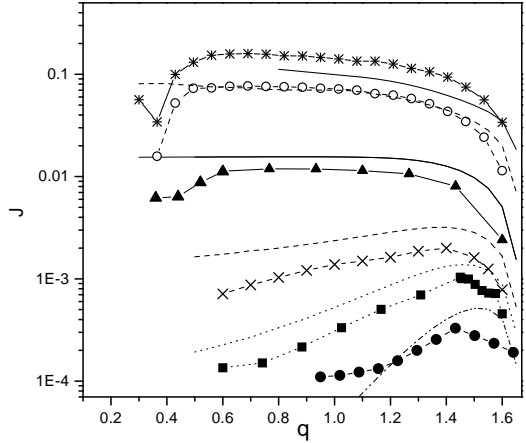}
\caption{Current as a function of $q$ for fixed $\tau_{ng}=100/(2
\pi)$ and different fixed values of $D_{ng}$. The lines with
symbols corresponds to simulations and the lines without symbols
to the adiabatic theory. From top to bottom, the curves are for
$D_{ng}=20$ (solid line for theory and solid line with stars for
simulations); $D_{ng}=5$ (dashed line for theory and dashed line
with hollow circles for simulations); $D_{ng}=1$ (solid line for
theory and solid line with triangles for simulations);
$D_{ng}=0.35$ (dashed line for theory and dashed line with crosses
for simulations); $D_{ng}=0.2$ (dotted line for theory and dotted
line with squares for simulations); and $D_{ng}=0.1$ (dash-dot-dot
line for theory and dash-dot-dot line with solid circles for
simulations). All calculations are for $T=0.5$ and $F=0.1$.}
\end{figure}

Finally, in order to analyze the relevance of the long tails of
the Non Gaussian distributions for $q>1$ in determining the value
of the current, we have done some special calculations. We have
simulated the dynamics of the Brownian motor forced by a different
-but related- non Gaussian noise. In Eq. (\ref{xpto}), instead of
$\eta(t)$ we consider the precess $\eta_{u}(t)$ which is defined
as $\eta_{u}(t)=\lambda_u(t) \eta(t)$, where $\lambda_u(t)$ is $1$
if $|\eta(t)|<u$, and $\lambda_u(t)=0$  if $|\eta(t)|>u$. Here,
$u>0$ is a parameter that plays the role of a threshold for the
non Gaussian noise, and indicates the maximum value of the noise
that can be ``feel'' by the particle. For $u \to \infty$ the
process $\eta_u(t)$ converges to $\eta(t)$. Note that, in
practice, in the simulations we calculate the complete evolution
for $\eta(t)$ as in the normal case, but we change the way in
which this noise couples to the Brownian particle.

In Fig.4.a we show results for the current as function of $u$ for
different values of $q$. It is apparent that the threshold $u$
needed to obtain the asymptotic value of the current
(corresponding to the one obtained with the process $\eta(t)$)
increases with $q$. This means that the tails of the distribution
are relevant up to higher values of $\eta$ as $q$ is increased. In
Fig. 4.b we plot, as a function of $q$, the value of the threshold
$u_c$ at which the asymptotic value of the current is reached with
an error lower than $5 \%$.

\begin{figure}
\centering
\includegraphics[width=10cm]{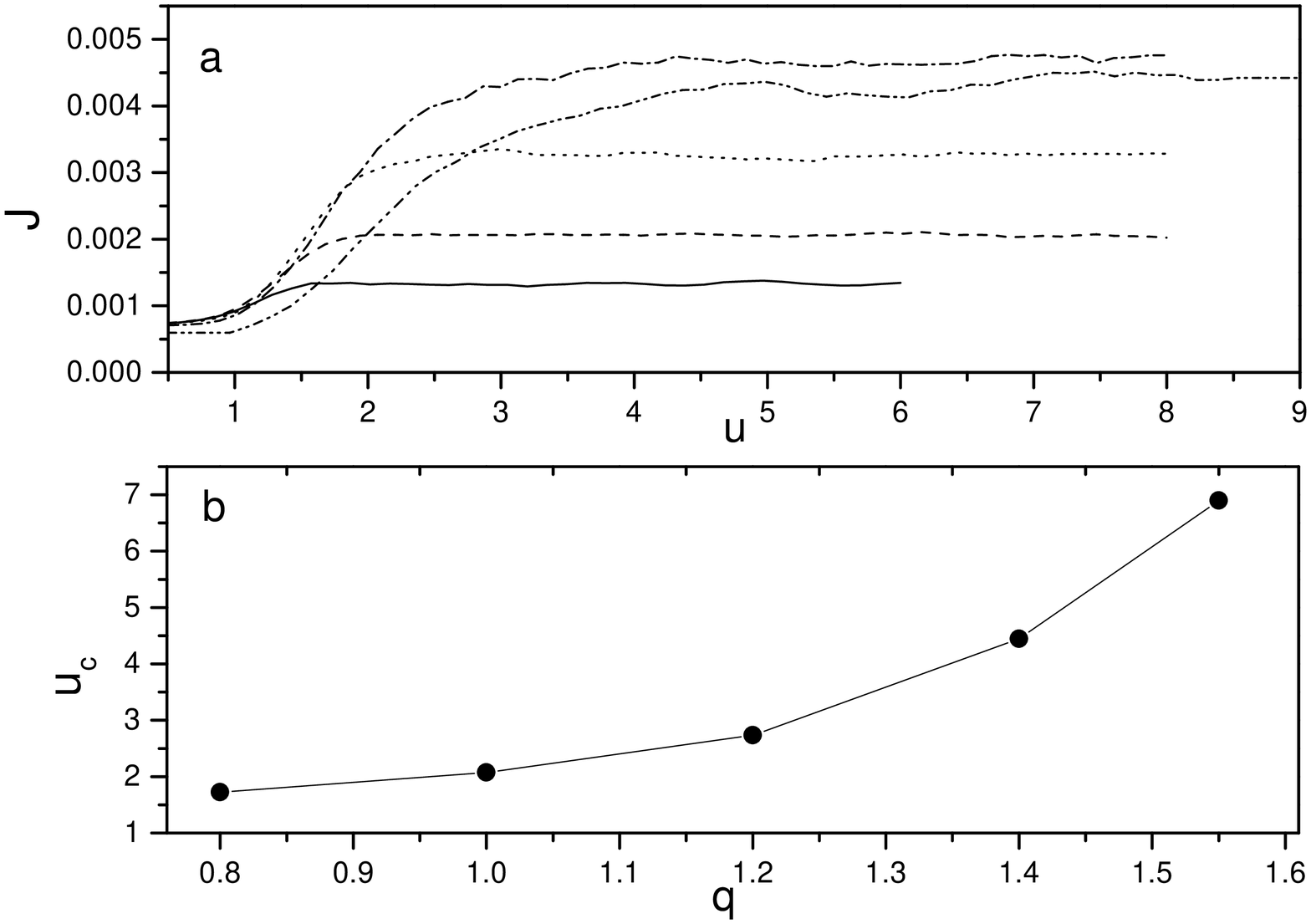}
\caption{(a) Current as a function of the threshold $u$ for
different values of $q$: Solid line for $q=8$, dashed line for
$q=1.$, dotted line for $q=1.2$, dash-dotted line for $q=1.4$, and
dash--dot--dotted line for $q=1.55$.}
\end{figure}

\section{Conclusions}

We have here further extended the study of the effect of non
Gaussian noises on the behavior and transport properties of
Brownian motors initiated in \cite{qruidoZ}. We have focused on
two aspects: (a) the adiabatic approximation (valid for a high
correlation time of the forcing) \cite{qruidoZ}, showing some
analytical results for the current $J$ as a function of $D_{ng}$
for different $q$; and (b) have analyzed the role played by the
long tails of the distribution, by analyzing the results of
truncated non Gaussian distributions.

By means of the adiabatic approximation as well as related
numerical simulations, we found that there is an ``optimal" value
of the parameter $q$, yielding the maximum possible value of the
current $J$ for a given value of $D_{ng}$. Also, that such an
optimum value can change from $q_o>1$ to $q_o<1$ for different
regions of values of $D_{ng}$. Regarding the analysis of truncated
non Gaussian distributions, we have shown the crucial role played
by the long tails in determining the enhancement of the Brownian
motor response.

These results complement those of \cite{qruidoZ} and supports the
finding of a strong influence of non Gaussian noises on the
response of Brownian ratchets, as was previously found for other
noise induced phenomena.

{\bf Acknowledgments:} HSW acknowledges the partial support from
ANPCyT, Argentine, and thanks to the European Commission for the
award of a {\it Marie Curie Chair} at the Universidad de
Cantabria, Spain.


\begin{thebibliography}{99}
\bibitem{comFP} In doing this, we use a piecewice linear
approximation for the ratchet potential which is
$V(x)=2.5-(5/1.1)x$ for $x<1.1$ and $V(x)=-2.5+[5/(2 \pi
-1.1)](x-1.1)$ for $1.1 < x < 2 \pi$. We solve the Fokker--Planck
equation for $0 \le x \le 2 \pi$ with periodic boundary condition
asking for continuity of the distribution and the current at
$x=1.1$.
\end{thebibliography}
\end{document}